\documentclass[aip,apl,reprint]{revtex4-1}
\renewcommand{\vec}[1]{\mathbf{#1}} 
\usepackage[]{graphicx}
\usepackage{graphics}
\usepackage{amsmath}
\usepackage{amsfonts}
\usepackage{amssymb}
\usepackage{wasysym} 
\usepackage{nicefrac} 
\usepackage{setspace}
\usepackage{color} 
\usepackage{nicefrac}
\usepackage{dcolumn}%
\usepackage{bm}%
\usepackage{parskip}

\renewcommand{\Re}{\operatorname{Re}}
\renewcommand{\Im}{\operatorname{Im}}

\newcommand{\figref}[1]{Fig.~\ref{fig:#1}}
\newcommand{\Figref}[1]{Figure~\ref{fig:#1}}

\renewcommand{\eqref}[1]{Eq.~(\ref{eq:#1})}

\begin{document}

\setlength{\parskip}{0pt} 
\setlength{\parindent}{10pt}
\setlength{\abovedisplayskip}{0pt}
\setlength{\belowdisplayskip}{-2pt}
\setlength{\abovecaptionskip}{-3pt}
\setlength{\belowcaptionskip}{-16pt}

\preprint{AIP/123-QED}

\title{Near-field refrigeration and tunable heat exchange through four-wave mixing}
 
\author{Chinmay Khandekar}
\email{cck@princeton.edu}
\address{Department of Electrical Engineering, Princeton University, Princeton, NJ 08544, USA}
\author{Riccardo Messina}
\email{riccardo.messina@umontpellier.fr}
\address{L2C, Universit\'{e} de Montpellier, CNRS, Montpellier 34095, France}
\address{Department of Electrical Engineering, Princeton University, Princeton, NJ 08544, USA}
\address{Laboratoire Charles Fabry, UMR 8501, Institut d'Optique, CNRS, Universit\'{e} Paris-Saclay, Palaiseau Cedex 91127, France}
\author{Alejandro W. Rodriguez}
\email{arod@princeton.edu}
\address{Department of Electrical Engineering, Princeton University, Princeton, NJ 08544, USA}

\begin{abstract}
  We modify and extend a recently proposed four-wave mixing scheme
  [Chinmay Khandekar and Alejandro Rodriguez, Optics Express, 25 (19),
    23164, 2017] for achieving near-field thermal upconversion and
  energy transfer, to demonstrate efficient thermal refrigeration at
  low intensities $\sim 10^{9}$W/m$^2$ over a wide range of gap
  sizes (from tens to hundreds of nanometers) and operational
  temperatures (from tens to hundreds of Kelvins). We further exploit
  the scheme to achieve magnitude and directional tunability of
  near-field heat exchange between bodies held at different
  temperatures.
\end{abstract}

\pacs{}

\keywords{}

\maketitle

Near field radiative heat
exchange~\cite{loomis1994theory,joulain2005surface,
  volokitin2007near,basu2009review,song2015near} is important for
several emerging applications and technologies, from energy
conversion~\cite{narayanaswamy2003surface,laroche2006near} to
nanoscale heat management and cooling.~\cite{cahill2014nanoscale,
  cahill2003nanoscale,shakouri2006nanoscale,guha2012near} This has
motivated recent efforts aimed at achieving active control of heat
transfer using gain media~\cite{ding2016active,khandekar2016giant} or
more generally, chemical potentials.~\cite{chen2015heat} Simultaneous
advances in nanofabrication have also made it possible to confine
light to small volumes and over long
timescales~\cite{soljavcic2004enhancement,armani2003ultra}, allowing
otherwise weak optical nonlinearities to modify even low-power
phenomena like thermal radiation~\cite{soo2016fluctuational,
  soo2017fluctuational,khandekar2015radiative,khandekar2015thermal}. We
recently proposed an alternative mechanism for controlling heat
exchange~\cite{khandekar2017near} that exploits nonlinear four-wave
mixing to extract and upconvert ``thermal energy" trapped in the near
field of a planar body unto another, from mid- to near-infrared
wavelengths~\cite{khandekar2017near}. In particular, we showed that
the combination of resonantly enhanced optical nonlinearities and
large density of states associated with tightly confined surface
plasmon/phonon--polariton (SPP) resonances enables high-efficiency
four-wave mixing in planar materials separated by nanoscale gaps,
resulting in order $10^5\,$W/m$^2$ upconversion rates induced by
externally incident mid-infrared light of moderate intensities, on the
order of $10^{12}$W/m$^2$.

\begin{figure}[!t]
\centering
\includegraphics[width=0.9\linewidth]{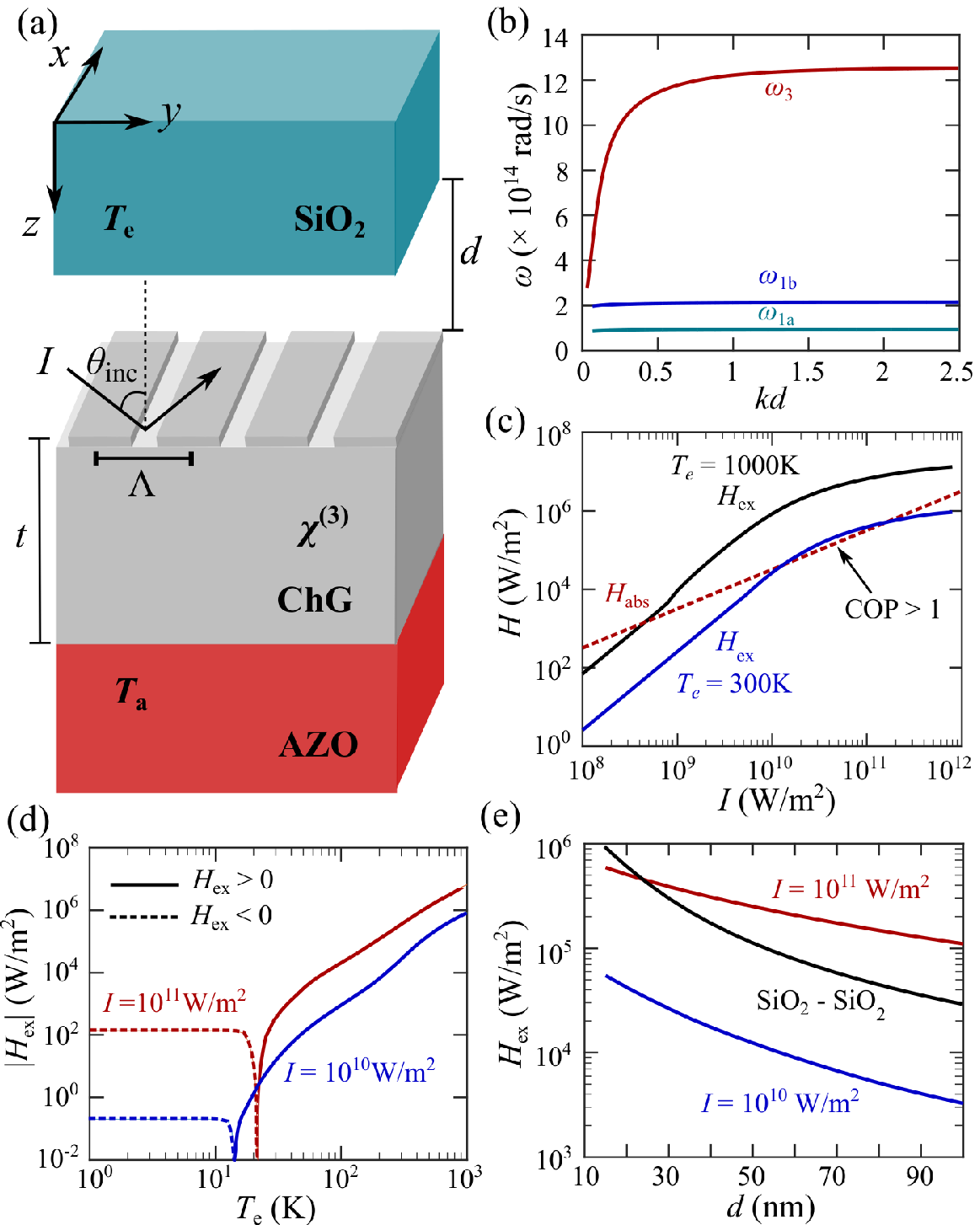}
\caption{(a) SiO$_2$ glass held at temperature $T_e$ is separated by a
  vacuum gap $d$ from an AZO slab at temperature $T_a$ that is coated
  with a ChG thin grating of thickness $t$ and period $\Lambda$. An
  external, monochromatic drive of intensity $I$ incident at angle
  $\theta_{\text{inc}}$ couples to a leaky mode of frequency
  $\omega_2$ in the ChG grating that mediates four-wave mixing between
  SiO$_2$ ($\sim \omega_{1a},\omega_{1b}$) and AZO ($\sim\omega_3$)
  SPPs. For fixed $d=30\,$nm, $t=100\,$nm, $T_a=300\,$K, (b) shows the
  associated SPP dispersions while (c) shows the extracted power
  $H_\text{ex}$ from SiO$_2$ at two different temperatures $T_e$
  (solid), along with the absorbed power $H_\text{abs}$ in the
  absorber (dashed), as a function of $I$.  (d) illustrates the
  dependence of $H_\text{ex}$ on $T_e$ for two values of drive
  intensities $I$, showing a reversal from positive $H_\text{ex} > 0$
  (solid lines) to negative $H_\text{ex} < 0$ (dashed lines)
  extraction at low temperatures, while (e) shows the dependence on
  vacuum gap $d$ for the same intensities but fixed $T_e=300\,$K. Also
  shown for comparison in the same plot is the flux rate between two
  semi-infinite SiO$_2$ plates held at temperatures $T_e=300\,$K and
  $T_a=0\,$K, as a function of $d$.}
\label{fig:fig1}
\end{figure}
\begin{figure}[!t]
\centering \includegraphics[width=0.9\linewidth]{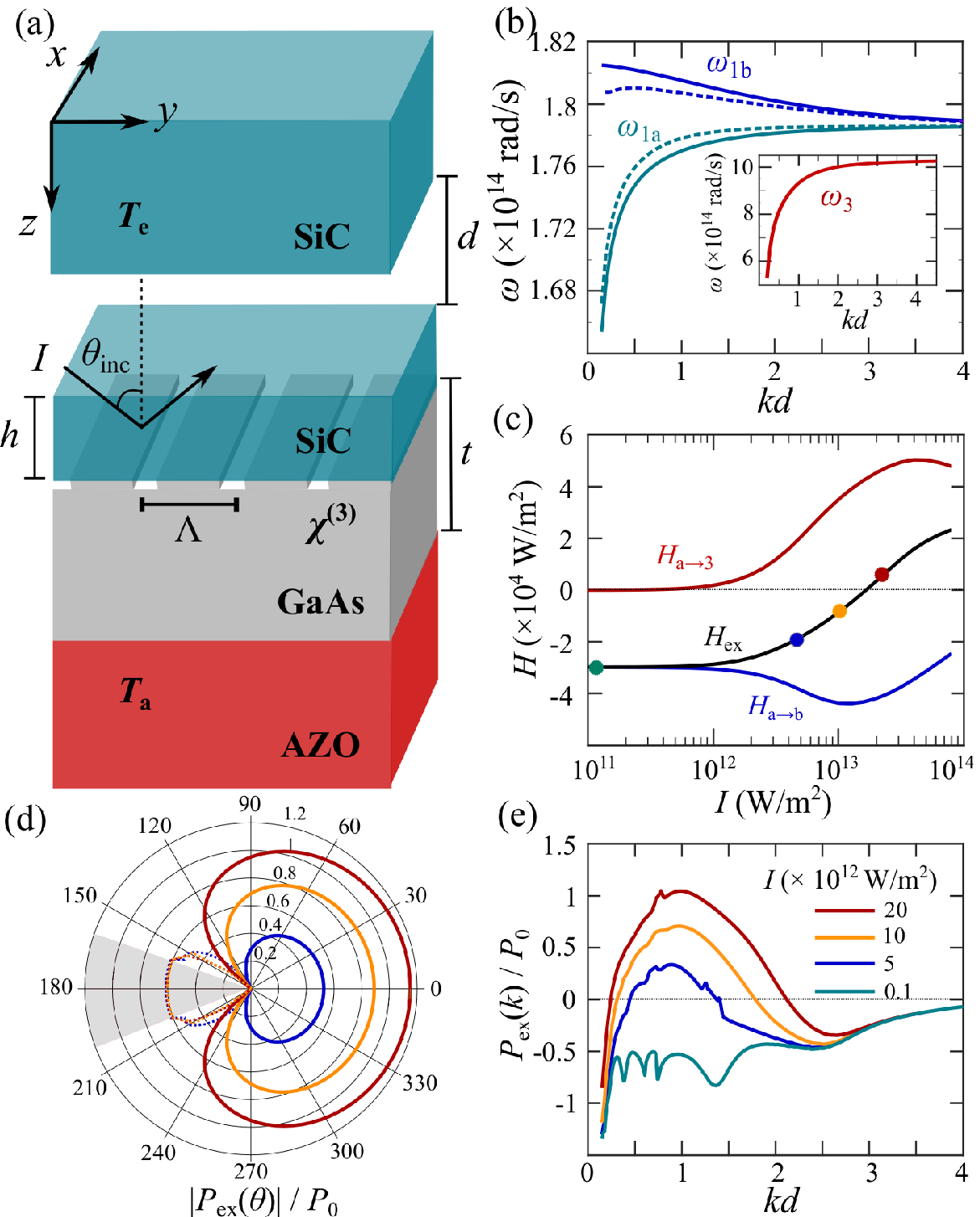}
\caption{(a) SiC emitter separated by a $d=50$nm vacuum gap from an
  AZO absorber that is coated with a composite layer consisting of a
  SiC film (thickness $h=20$nm) on top of a nonlinear GaAs film
  (thickness $t=100$nm). The thin periodic grating couples externally
  incident light at $\theta_\text{inc}$ and intensity $I$ to a
  mediator mode at $\omega_2$ in GaAs, mediating four-wave mixing
  between various SPP resonances, with dispersions shown in (b),
  localized to both AZO ($\omega_3$) and SiC
  ($\omega_{1a},\omega_{1b}$) interfaces, where solid (dashed) lines
  denoting perturbed (unperturbed) modes (see text). Assuming a
  temperature differential, $T_e=300$~K and $T_a=400$~K, (c) shows the
  net extraction power from the SiC slab across the gap $H_\text{ex}$,
  along with the individual nonlinear $H_{a \to 3}$ and linear $H_{a
    \to b}$ contributions, illustrating the possibility of tunable
  extraction (in both magnitude and direction) with respect to drive
  intensity $I$. (d) and (e) illustrate the underlying contribution of
  individual resonances, classified by their angular and wavevector
  states, showing the per-angle $P_\text{ex}(k,\theta)$ (at $kd=0.5$)
  and angle-integrated frequency-integrated flux $P_\text{ex}(k)$
  (with respect to $k$), respectively, at multiple $I$, marked by
  circles in (c). The flux rates are normalized by
  $P_0=2\gamma_1\Theta(\omega_1,T_e)$, where $\gamma_1=4.45\times
  10^{14}\,$rad/s and $\omega_1=1.8\times 10^{14}\,$rad/s.}
\label{fig:fig2}
\end{figure}

In this letter, we show that a similar four-wave mixing scheme can be
exploited to achieve thermal refrigeration and tunable heat
exchange. We begin by exploring the planar configuration shown in
\figref{fig1}(a), comprising an emitter held at temperature $T_e$ and
supporting mid-infrared SPP resonances around frequency $\omega_1$
which is separated by a vacuum gap from an absorber held at
temperature $T_a$ and supporting near-infrared SPPs around
$\omega_3$. The absorber is coated with a thin $\chi^{(3)}$ nonlinear
film supporting a mediator resonance at $\omega_2 \sim
(\omega_3-\omega_1)/2$ which couples to externally incident light by
way of a grating. The mediating mode facilitates resonant four-wave
mixing ($\omega_1+2\omega_2 = \omega_3$) between the SPP resonances,
resulting in cooling of the emitter by way of upconversion and energy
transfer across the gap. As shown below, in contrast to passive
radiative cooling mechanisms requiring large temperature differentials
$T_e \gg T_a$, nonlinear upconversion allows thermal energy extraction
under zero or even negative differentials ($T_e < T_a$), constrained
only by photon-number conservation~\cite{khandekar2017near}. This in
turn enables thermal refrigeration, where thermal energy is made to
flow from a low to high temperature body (a reversed heat engine) when
the system is driven by external light, which provides the work
required for energy transfer. The first part of this letter is devoted
to a detailed analysis of such a refrigeration scheme, illustrating
not only the various design criteria but also operating regimes needed
to achieve high-efficiency refrigeration, including temperature range
($T_e \sim 10$--$1000\,$K) and gap sizes. In the second part, we
extend the analysis to consider a more complicated system, depicted in
\figref{fig2}(a), where we introduce an additional thin film on top of
the nonlinear medium for the purpose of enabling appreciable heat
exchange under zero external drive but finite temperature
differentials $T_e \neq T_a$, otherwise absent due to the large SPP
frequency mismatch between the emitter and absorber. This channel can
thus compete with nonlinear energy upconversion to enable tunable heat
flow (in both magnitude and direction) with respect to the incident
drive power.


A significant refinement in this paper with respect to our earlier
work~\cite{khandekar2017near} is the substitution of lossy plasmonic
resonances in favor of low-loss dielectric leaky modes in the
nonlinear medium. While the latter are less localized than the former,
they exhibit longer radiative and absorptive lifetimes and thus result
in significantly lower power requirements, on the order of
$10^{9}\,$W/m$^2$ as opposed to $10^{12}\,$W/m$^2$, while also
mitigating pump-induced heating. While the choice of transparent
materials around the pump wavelength $\omega_2$ mitigates heating
introduced by the drive, in practice we expect that efficient thermal
cooling will require a vacuum gap (not considered
before~\cite{khandekar2017near}) in order to further limit conductive
transfer stemming from spurious heating.  Our theoretical analysis is
based on a coupled-mode theory
framework,~\cite{haus1984waves,fan2003temporal,ruan2009temporal}
previously exploited to analyze heat transfer in linear
media~\cite{chalabi2014ab,lin2017application,iizuka2016temporal} and
more recently generalized to consider a broad class of weakly
nonlinear resonant processes~\cite{joannopoulos2011photonic,
  rodriguez2007chi}, that provides general operating conditions and
quantitative predictions while allowing us to avoid otherwise
cumbersome calculations based on nonlinear fluctuational
electrodynamics.~\cite{soo2016fluctuational} Finally, we note that our
predictions extend recent work in the area of non-contact
refrigeration~\cite{ding2016active,liu2016high,chen2017high} and
dynamically tunable heat
exchange,~\cite{ito2017dynamic,khandekar2016giant} and has analogies
with more established thermoelectric cooling
schemes.~\cite{riffat2004improving}

\emph{Thermal refrigeration.---}We first consider the planar system
shown in \figref{fig1}(a), comprising a silica (SiO$_2$) emitter
separated by a vacuum gap $d$ from an aluminum-doped zinc oxide (AZO)
absorber. The associated dielectric properties are obtained from
various references.~\cite{palik1998handbook,pirozhenko2008influence,
  kim2013plasmonic} The nonlinear medium is a chalcogenide (ChG) thin
film of material composition As$_2$S$_3$, thickness $t$, permittivity
$\varepsilon_2=6.25$, and isotropic Kerr coefficient
$\chi^{(3)}=10^{-17}\,\text{m}^2/\text{V}^2$.~\cite{aio1978refractive,
  yan2011third,zakery2003optical,harbold2002highly} (Note that we
assume an isotropic Kerr coefficient,
$\chi_{xxxx}=3\chi_{xxyy}=3\chi_{xyxy}=\chi^{(3)}$, purely for
computational and conceptual convenience, but that more generally the
nature of the relevant tensor components will depend on growth and
material considerations.\cite{yan2011third, boyd2003nonlinear}) The
$p$-polarized SPP resonances in this configuration are characterized
by their conserved transverse momenta $\vec{k}$ and described by mode
profiles of the form $E_{l}(z)e^{i\vec{k}.\vec{x}_\parallel}$, where
$l \in {x,y,z}$ and $\vec{x}_\parallel$ is the transverse
position. \Figref{fig1}(b) shows multiple mode dispersions $\omega(k)$
arising in the above configuration, for a choice of $d=30$nm,
illustrating two branches of SPPs localized at the SiO$_2$ interface
of frequencies $\omega_{1a} \sim 2\times 10^{14}\,$rad/s and
$\omega_{1b} \sim 0.8 \times 10^{14}\,$rad/s, along with a single SPP
branch localized at the AZO interface of frequency $\omega_3 \sim
12\times 10^{14}\,$rad/s. Also present (not shown) is a separate
mediator resonance that propagates primarily within the ChG film with
frequency $\omega_2 \approx 5\times 10^{14}\,$rad/s and wavevector
$\vec{k}_2 = k_2 \hat{y}$. This mediator mode can couple to externally
incident light at $\omega_2$ and angle $\theta_{\text{inc}}$ by way of
a thin, first-order diffraction grating of period $\Lambda$, designed
to satisfy $\dfrac{\omega_2}{c}\sin \theta_{\text{inc}} +
\dfrac{2\pi}{\Lambda}= k_{2}$. Note that the thin size of the grating
($\lesssim 5$nm) and large frequency mismatch between the mediator and
SPP resonances combine such that the grating has a negligible impact
on the dispersions and resonances of the
slabs.~\cite{sharon1996narrow} Furthermore, four-wave mixing between
SPPs is only possible under the momentum-matching condition,
$\vec{k}_1 + 2\vec{k}_2 = \vec{k}_3$,~\cite{khandekar2017near} thus
ensuring that only a single emitter mode at $\vec{k}_1$ couples
exclusively to an absorber mode at $\vec{k}_3$. In particular, given a
set ($\vec{k}_1,\vec{k}_3$) of momentum-matched modes, the
upconversion rates can be computed using the following coupled mode
equations:
\begin{align} \label{eq:eqa1a3}
  \dot{a}_{1\alpha} &= (i\omega_{1\alpha}-\gamma_{1\alpha})a_{1\alpha}
  -i\kappa_{1\alpha}e^{-2i\omega_2 t}a_3 - i\kappa_{l}a_{1\beta} \nonumber \\
  &+ \sqrt{2\gamma_{1\alpha}}\xi_{1\alpha}, \quad \alpha,\beta \in
  \{a,b\}, \quad \alpha\neq\beta \\ \dot{a}_3 &= (i\omega_3
  -\gamma_3)a_3 - \sum_{\alpha=a,b}
  i\frac{\omega_3}{\omega_{1\alpha}}\kappa_{1\alpha}^* e^{2i\omega_2
    t}a_{1\alpha}+ \sqrt{2\gamma_3}\xi_3, \nonumber
\end{align}
where $a_j$ denotes the amplitude of mode $j \in [1a, 1b, 2, 3]$,
normalized such that $|a_j|^2$ is the corresponding mode energy and
$\xi_j$ represent thermal noise sources with thermodynamic
correlations $\langle \xi_j^* (\omega)\xi_j(\omega') \rangle =
\Theta(\omega,T_k)\delta(\omega-\omega')$, where
$\Theta(\omega,T_j)=\hbar\omega/[\text{exp}(\hbar\omega/k_BT_j)-1]$
denotes the Planck distribution corresponding to a local bath
temperature $T_j$.~\cite{karatzas2012brownian} The mode frequencies
$\omega_j(k_j)$ and associated decay rates $\gamma_{j}$ are obtained
from the complex, eigenfrequency solutions of Maxwell's equations,
while the nonlinear coupling coefficients $\kappa_{1\alpha}$
($\alpha=a,b$) describing four-wave mixing are obtained via
perturbation theory~\cite{joannopoulos2011photonic, rodriguez2007chi}
and depend on a complicated, spatial overlap of the linear profiles
within the nonlinear medium:~\cite{khandekar2017near}
\begin{widetext}
\begin{align}
\kappa_{1\alpha}(\vec{k}_1,\vec{k}_3) = \frac{\omega_{1\alpha}\gamma_c
  I\int dz
  \chi_{ijkl}^{(3)}e^{i(\vec{k}_3-\vec{k}_1-2k_{2}\hat{y}).\vec{x}_\parallel}
  E_{1\alpha,i}(\omega_{1\alpha},z)
  E_{2,j}(\omega_2,z)E_{2,k}(\omega_2,z)E_{3,l}^*(\omega_3,z)}{
  \epsilon_0\gamma_{2t}^2 (\int dz \frac{\partial
    \epsilon\omega}{\partial \omega}|E_2(\omega_2,z)|^2)(\int dz
  \frac{\partial \epsilon\omega}{\partial
    \omega}|E_{1\alpha}(\omega_{1\alpha},z)|^2)^{1/2}(\int dz
  \frac{\partial \epsilon\omega}{\partial
    \omega}|E_3(\omega_3,z)|^2)^{1/2}}.
\label{eq:kappa}
\end{align}
\end{widetext}
Here $I$ is the drive intensity and
$\gamma_{2t}=\gamma_{2}+\gamma_{2c}$ is the overall loss rate of the
mediator resonance, which includes both dissipative $\gamma_{2}$ and
radiative decay $\gamma_{2c}$ (induced by the periodic grating). Note
that the momentum-matching condition of nonzero coupling follows by
inspection of the phase factor $e^{i(\vec{k}_3-\vec{k}_1-2k_2
  \hat{y})}$, allowing us to simplify
$\kappa_{1\alpha}(\vec{k}_1,\vec{k}_3)\rightarrow
\kappa_{1\alpha}(k,\theta)$, where $k=|\vec{k}_1|$ and $\theta$ is the
angle between $\vec{k}_1$ and the wavevector of the mediator mode
(parallel to the $y$ axis). Also included in the coupled-mode
equations is the possibility of finite linear coupling $\kappa_l$
between the two emitter SPPs, which is negligible in the current
configuration due to the large discrepancy between $\omega_{1a}$ and
$\omega_{1b}$ but turns out to be of critical importance in the
configuration of \figref{fig2}(b).

From these coupled-mode equations, one can find the various energy
transfer rates corresponding to a given set of modes ($k,\theta$) by
considering the overall energy loss rate associated with each
mode,~\cite{haus1984waves} leading to simple expressions for the
thermal extraction $P_{\alpha\to 3}=2\langle \Im[\kappa_{1\alpha}^*
\text{exp}(2i\omega_2 t)a_3^* a_{1\alpha}] \rangle$ and linear heat
transfer $P_{a\to b}=2\langle \Im[\kappa_l^* a_{1a}a_{1b}^*]\rangle$
rates, along with associated power spectral
densities.~\cite{khandekar2017near} The net flux rates
$H_{\alpha\to\beta}$ are then given by:
\begin{align}
  H_{\alpha \to \beta} = \int_0^{\infty}\frac{k dk }{2\pi}
  \overbrace{\int_0^{\infty}\frac{d\theta}{2\pi}
    \underbrace{\int_0^{\infty}\frac{d\omega}{2\pi}
      P_{\alpha\to\beta}(\omega,k,\theta)}_{P_{\alpha\to\beta}(k,\theta)}}^{P_{\alpha\to\beta}(k)},
\end{align}
where $\alpha,\beta \in \{a,b,3\}$ labels the particular flux channel.

To provide a proof-of-concept demonstration of thermal refrigeration,
we first consider typical geometric and operating parameters, with
$d=30$nm, $t=100$nm, $\Lambda=2\mu$m,
$\theta_{\text{inc}}=45^{\circ}$, and $T_a=300K$. \Figref{fig1}(c)
shows the net thermal extraction rate $H_\text{ex} =H_{a \to 3}+H_{b
  \to 3}$ corresponding to two different emitter temperatures,
$T_e=300\,$K (blue curve) and $T_e=1000\,$K (black curve), as a
function of drive intensity $I$. Evidently, large flux rates
$H_{\text{ex}} \sim 10^5\,$W/m$^2$ are achievable with moderate drive
intensities $I \sim 10^{9}\,$W/m$^2$, illustrating over three
orders of magnitude improvements in power efficiency (reduced
intensity requirements) over earlier
configurations~\cite{khandekar2017near} based on lossy plasmonic
mediator resonances. Note that the transparency of the emitter at pump
wavelength as well as the presence of vacuum gap and large SPP
frequency mismatch imply that conductive or radiative heating of the
emitter due to the pump is negligible compared to heat extraction
leading to its cooling. The efficiency of such reversed heat engine
(refrigeration scheme) is given by a coefficient of
performance~\cite{riffat2004improving} (COP), defined as the ratio of
thermal energy extracted to power that is lost to pump-induced
heating. It follows from coupled mode equations that the absorbed pump
intensity at $\omega_2$ is given by
$H_{\text{abs}}=4\gamma_{2}\gamma_{2c}I/\gamma_{2t}^2$. We choose the
radiative or coupling rate of the mediator mode to be $\gamma_{2c} =
10^{-5} \omega_2$, a very reasonable estimate based on extensive
theoretical~\cite{rosenblatt1997resonant,chang2012high, levy2000very}
and experimental work on similar thin
gratings.~\cite{sharon1996narrow} The dissipation rate $\gamma_{2}
\approx \Im\{\epsilon_{\text{m}}\}\omega_2/
2\Re\{\epsilon_{\text{m}}\}$ where $\epsilon_{\text{m}}$ is the
complex permittivity of the nonlinear medium, is obtained from
perturbation theory and agrees with the exact complex eigenfrequency
solution. With $\Im\epsilon_{\text{m}} \approx 10^{-10}$ (obtained
from extrapolating available data~\cite{adam2014chalcogenide}), it
follows that $\gamma_{2} \ll \gamma_{2c}$ and as shown by red curve in
\figref{fig1}(c), the absorbed power $H_{\text{abs}}\approx 4\gamma_2
I/\gamma_{2c}$ by the ultra-low loss resonance is smaller than the
heat extraction rates leading to COP $\gg 1$ over a varying range of
intensities. While such ultra-low loss resonances have been explored
extensively in different context~\cite{yang2018bridging}, they play an
important role here in minimizing the unnecessary power dissipation
and enhancing the refrigeration efficiency (COP). While ideally, COP
$>1$ is within reach, various non-idealities such as fabrication
imperfections and spurious material loss rates may lead to effectively
larger dissipation rates in actual experiments and potentially smaller
values of COP $\approx 10^{-2}$. We note that these are realistic
efficiencies of all such solid-state refrigeration
schemes~\cite{chen2017high,riffat2004improving} and are acceptable
given the reliability (no moving parts) and ease of on-chip
implementation in comparison to other gas based refrigeration methods.

Yet another important figure of merit is the range of operating
temperatures over which it is possible to cool the emitter
(independently of efficiencies). Along this vein, \figref{fig1}(d)
shows $H_\text{ex}$ as a function of $T_e$ for multiple values of $I$,
illustrating a change in the sign of the flux from positive (solid) to
negative (dashed) as $T_e$ decreases past a typical transition
temperature $T_e \sim 10\,$K. It follows that under ideal conditions
under which $T_a=300\,$K is held fixed, heat can be extracted from the
emitter until it is cooled down to temperatures on the order of tens
of Kelvin.  Moreover, while we have chosen so far to focus on
configurations involving very small vacuum gaps $d=30\,$nm, as shown
in \figref{fig1}(e), significant flux rates can nevertheless be
achieved at larger separations $d\sim 100\,$nm. This is an important
consideration for current state of the art experiments exploring
near-field heat transfer in planar
geometries~\cite{song2015near}. Interestingly, we find that the
complicated dependence of the spectral flux rate on gap and drive
intensity, quantified by the coupling coefficient $\kappa(k,\theta)$,
leads to a modified relationship between the net flux and gap size
compared to the typical $\sim 1/d^2$ dependence associated with linear
heat exchange. Comparing $H_\text{ex}$ in the particular case of
$I=10^{11}$~W/m$^2$ and equal $T_e=T_a=300\,$K (solid red line) to
the flux rate between two SiO$_2$ plates held at a large temperature
differential, $T_e=300\,$K and $T_a=0\,$K, but separated by the same
gap sizes (black line), one finds not only significantly larger
extraction rates but also a slower polynomial decay in the former. We
finally remark that apart from the efficiency comparable to other
solid-state refrigeration
methods~\cite{chen2017high,riffat2004improving}, the design
flexibility and wide range of temperature regimes achievable through
this scheme could prove a viable alternative depending on the
application.

\emph{Tunable heat exchange:} We now consider a slightly modified
configuration, depicted schematically in \figref{fig2}(a), to
illustrate the possibility of exploiting four-wave mixing as a means
of achieving tunable heat exchange. The modified configuration
consists of a silicon carbide (SiC) emitter separated by vacuum from a
aluminum-doped zinc oxide (AZO) absorber. Resting on the absorber is a
composite layer consisting of an additional SiC thin film of thickness
$h=20$nm on top of a nonlinear gallium arsenide (GaAs) film, with
nonlinear $\chi^{(3)}=10^{-18}\,$m$^2/\text{V}^2$ and dielectric
properties taken from various
references.~\cite{palik1998handbook,pirozhenko2008influence,
  kim2013plasmonic,blakemore1982semiconducting} The presence of the
additional SiC thin film results in three branches of SPPs, out of
which only two, depicted as solid lines in \figref{fig2}(b),
correspond to SPPs localized along the SiC--vacuum interfaces. In
contrast to the previous configuration, however, these SPPs have
non-negligible linear coupling ($\kappa_l\neq 0$) due to similar
resonance frequencies and can therefore contribute significant linear
heat exchange between the emitter and absorber. In order to
incorporate both nonlinear upconversion and linear heat exchange, we
obtain the unperturbed frequencies, mode profiles, and linear coupling
rates $\kappa_l$ of the SPPs by fitting the full linear fluctuational
electrodynamics calculations to the coupled-mode equations. The dashed
(solid) lines in \figref{fig2}(b) depict the unperturbed (perturbed)
resonance frequencies, $\omega_{1a} \sim \omega_{1b} \sim 1.8\times
10^{14}$rad/s, obtained using the fitting procedure. In addition, the
system supports a SPP branch around frequency $\omega_3 \sim 10\times
10^{14}\,$rad/s that is localized along the AZO interface, shown as an
inset (red line). Like before, a thin grating of period
$\Lambda=1.78\,\mu$m is used to couple incident light at angle
$\theta_{\text{inc}}=45^{\circ}$ to a mediator resonance in the GaAs
film of frequency $\omega_2\approx 4\times 10^{14}\,$rad/s and
wavevector $\vec{k}_2 = k_2 \hat{y}$.

To provide a proof-of-concept demonstration of tunable heat exchange,
we consider a typical set of parameters, corresponding to $d=50\,$nm,
$h=20\,$nm, $t=100\,$nm, and $T_e=300\,$K$ < T_a=400\,$K, which in the
absence of the drive nevertheless leads to a net heat exchange across
the gap directed toward the emitter. \Figref{fig2}(c) shows the net
$H_\text{ex} = H_{a\to b} + H_{a \to 3}$ (black line) along with
individual $H_{a \to b}$ (blue line) and $H_{a \to 3}$ (red line)
extraction rates, as a function of drive intensity $I$. Evidently, at
low drive intensities $I\ll 10^{12}$~W/m$^2$, the net extraction rate
$H_\text{ex}$ is dominated by the linear heat exchange between the SiC
resonances ($H_{a\to b}<0$), becoming gradually larger due to
nonlinear extraction ($H_{a\to 3}>0$) with increasing $I$, with the
reversal in heat flow across the gap occurring at $I\gtrsim
10^{13}$~W/m$^2$. Note that while intuitively one might expect a
decreasing amount of linear heat flow with increasing nonlinear
upconversion, we observe a non-monotonic trend in $H_{a\to b}$ with
increasing $I$ that demonstrates instead an increase in linear flow at
low intensities. Such a non-trivial interplay between the two (linear
and nonlinear) processes originates from a shift in the SiC mode
frequencies that ends up enhancing the otherwise sub-optimal (due to
the slight frequency mismatch) linear heat flow. While the net flux
rates contain contributions from a wide set of SPPs, characterized by
$(k,\theta)$, the underlying behavior of flux rates for these modes
can be analyzed by inspection of the frequency-integrated spectral
flux $P_\text{ex}(k,\theta)$. For illustration, \figref{fig2}(d) shows
the angular dependence of the flux rate at a fixed $k d=0.5$, with
\figref{fig2}(e) showing the underlying angle-averaged spectrum with
respect to $kd$ at multiple drive intensities, corresponding to the
points marked by circles in \figref{fig2}(c). For convenience, we
normalize these flux rates by $P_0=2\gamma_1\Theta(\omega_1,T_e)$, or
the thermal power available to a single SPP, for typical values of
$\gamma_1=4.45\times 10^{11}\,$rad/s and $\omega_1=1.78\times
10^{14}\,$rad/s. As illustrated in \figref{fig2}(d), both the
magnitude and direction of the flux rate depend on the angle $\theta$
and intensity $I$, with the latter eventually resulting in large,
positive angle-averaged flux rates at larger $I$. Note that there
exists a range of modes, corresponding to highly acute angles (grey
region), for which the momentum-matching condition
$\vec{k}_3=\vec{k}_1+2\vec{k}_2$ can never be satisfied and for which
there is no nonlinear upconversion. Finally, \figref{fig2}(e) shows
the growing contribution and role of modes satisfying
momentum-matching in the net exchange, allowing nonlinear upconversion
to overwhelm the linear heat flow with increasing $I$.

\emph{Concluding remarks.---}We demonstrated a four-wave mixing scheme
for active near-field heat extraction. This approach not only enables
efficient nanoscale thermal refrigeration at very low temperatures
$\sim 10\,$K and low input intensities $I \sim 10^{9}$~W/m$^2$, but
also active control of both the magnitude and direction of heat flow
across vacuum gaps. While the systems explored in this work represent
only a proof of concept, we are confident that other geometries and
materials could result in further improvements.We note that the
coupled mode approach~\cite{khandekar2017near} employed above is valid
as long as the decay and coupling rates are much smaller than the
resonant frequencies of SPPs. It breaks down for large intensities $I
\gtrsim 10^{14}$~W/m$^2$ where not only coupling rates are very large
but also other considerations such as optical damage
threshold~\cite{naftaly2016silicon} become important. While coupled
mode theory circumvents the need to carry out full and repeated
calculations, further analysis of such nanoscale pump-thermal mixing
processes using nonlinear fluctuational
electrodynamics~\cite{soo2016fluctuational} remains a challenging and
interesting problem for future work.

\emph{Acknowledgments.---} This work was partially supported by the
National Science Foundation (DMR-1454836), Princeton Center for
Complex Materials with funding from NSF MRSEC program (DMR-1420541)
and Cornell Center for Materials Research with funding from the NSF
MRSEC program (DMR-1719875). 

\bibliographystyle{Riccardo} \bibliography{photon2}
\end{document}